\documentclass[conference]{IEEEtran}
\IEEEoverridecommandlockouts
\usepackage{tikz}
\usetikzlibrary{positioning, calc, fit, arrows.meta}

\usepackage{adjustbox}   
\usepackage{multirow}
\usepackage{cite}
\usepackage{amsmath,amssymb,amsfonts}
\usepackage{algorithmic}
\usepackage{subcaption}
\usepackage{siunitx}
\usepackage{textcomp}
\usepackage{graphicx}
\usepackage[ruled,vlined,linesnumbered]{algorithm2e} 
\SetAlCapFnt{\small}
\SetAlCapNameFnt{\small}
\usepackage{graphicx}
\usepackage{textcomp}
\usepackage{xcolor}
 \usepackage{booktabs} 
\DeclareUnicodeCharacter{03BB}{$\lambda$}
\DeclareUnicodeCharacter{0394}{$\Delta$}

\DeclareUnicodeCharacter{2009}{\,}  
\def\BibTeX{{\rm B\kern-.05em{\sc i\kern-.025em b}\kern-.08em
    T\kern-.1667em\lower.7ex\hbox{E}\kern-.125emX}}
\begin{document}

\title{Quantum Semi-Random Forests for Qubit-Efficient Recommender Systems}

\author{
\IEEEauthorblockN{Azadeh Alavi\IEEEauthorrefmark{1}, Fatemeh Kouchmeshki\IEEEauthorrefmark{2}, Abdolrahman Alavi\IEEEauthorrefmark{2}, Yongli Ren\IEEEauthorrefmark{1}, Jiayang Niu\IEEEauthorrefmark{1}}\\
\IEEEauthorblockA{\IEEEauthorrefmark{1}RMIT University, Melbourne, Australia\\
Email: \{azadeh.alavi, yongli.ren, s4068570\}@rmit.edu.au}
\IEEEauthorblockA{\IEEEauthorrefmark{2}Pattern Recognition Pty. Ltd., Melbourne, Australia\\
Email: \{admin@pr2aid.com, rahmanalavi1944@gmail.com\}}\\
\textit{\small{\textbf{\textit{First and second authors contributed equally to this work}}}}
}

\maketitle

\begin{abstract}
Modern recommenders describe each item with
hundreds of sparse semantic tags, yet most quantum pipelines still map one qubit per tag, demanding well beyond one hundred qubits, far out of reach for current noisy-intermediate-scale quantum (NISQ) devices and prone to deep, error-amplifying circuits. We close this gap with a three-stage hybrid machine learning algorithm that compresses tag profiles, optimizes feature
selection under a fixed qubit budget via QAOA, and scores
recommendations with a Quantum semi-Random Forest (QsRF) built on just five qubits, while performing similarly to the state-of-the-art methods. Leveraging SVD sketching and k-means, we learn a 1 000-atom dictionary ($>$97 \% variance), then solve a 20×20 QUBO via depth-3 QAOA to select 5 atoms. A 100-tree QsRF trained on these codes matches full-feature baselines on ICM-150/500.

\end{abstract}

\section{Introduction \& Related Work}
\label{sec:intro}

Modern content platforms such as \textit{Netflix}, \textit{Amazon} and
\textit{TikTok} describe every item with hundreds of sparse semantic
tags.  Identifying the smallest subset that preserves ranking quality is
NP–hard, and naïvely mapping one qubit to one tag quickly surpasses the
limits of Noisy-Intermediate-Scale-Quantum (NISQ) hardware.  To compress
this combinatorial explosion, recent hybrid pipelines formulate feature
selection as a \textbf{Q}uadratic \textbf{U}nconstrained \textbf{B}inary
\textbf{O}ptimisation (QUBO) and delegate the search to quantum
annealers \cite{niu2024caqubo,carletti2023pdqubo} or shallow
gate-based circuits \cite{mucke2023feature}.  Nevertheless, the entire
150-tag space in the smallest QuantumCLEF split is still embedded onto
hardware with limited connectivity, leading to deep, error-prone
circuits.

Historically, recommender systems relied on dense latent factor models
whose dimensions are hard to interpret.  In contrast, dictionary
learning provides compact and inherently sparse representations.
Early large-scale work by Mensch \textit{et al.} \cite{mensch2016scalable}
introduced a K–SVD-style factorisation that scales to terabyte user–item
matrices, while Kartoglu and Spratling \cite{kartoglu2018dictionary}
demonstrated that structured sparse coding can rival dense baselines in
both rating prediction and top-$N$ ranking with far fewer parameters.
Such results suggest that a well-chosen set of interpretable atoms can
replace raw tags without sacrificing accuracy—an observation we exploit
\emph{before} invoking quantum search.

Quantum feature selection has so far moved along two strands.  On the
one hand, mutual-information and counterfactual criteria have been
encoded directly into QUBO matrices and solved with either quantum
annealing or the Quantum Approximate Optimisation Algorithm
(QAOA) \cite{turati2022feature,nembrini2021qrec,carletti2023pdqubo}.
On the other hand, recursive variants such as RQAOA iteratively fix
strongly correlated variables to keep qubit counts manageable, albeit at
the cost of multiple quantum evaluations and heuristic
rounding \cite{bravyi2020rqaoa}.  Despite these innovations, published
studies still operate on only 10–20 features and remain bottlenecked by
the \emph{linear} growth in qubit demand.

Meanwhile, the QAOA landscape itself has matured from proof-of-concepts
to medium-scale demonstrations.  Harrigan \textit{et al.} executed
depth-2 QAOA on 23 superconducting qubits, validating theoretical
predictions but exposing depth-related locality
issues \cite{harrigan2021quantum}.  Follow-up work introduced
non-local mixers, parameter-transfer heuristics and hybrid decomposition
schemes, yet a definitive advantage over classical solvers is still
elusive.  Ferrari Dacrema \textit{et al.} systematically compared QPU,
hybrid, simulated-annealing and tabu-search solvers on up to 5 000
features and found quantum approaches merely \emph{competitive} once
problem sizes required decomposition \cite{ferraridacrema2022sigir}.
Consequently, qubit-efficient representations remain the critical
enabler for any near-term quantum recommender.

\subsection{Why Gate-based QAOA over Quantum Annealing?}
\label{sec:hardware}
Quantum annealers excel at native Chimera/Pegasus topologies but incur a
\emph{minor-embedding} overhead for fully-connected QUBOs: a single
logical variable is mapped to a ferromagnetically coupled \emph{chain}
of physical qubits whose expected length scales as
$\mathcal{O}(\sqrt{M})$ \cite{pelofske2023annealervsqaoa}.  For our
20×20 performance-driven QUBO this would raise the effective budget from
5 logical qubits to ${>}60$ physical qubits—even before error-correcting
replica chains are considered.  Chain breaks further degrade solution
quality and require post-processing heuristics
\cite{embedding2025minor}.

Gate-based devices avoid minor embedding: the five logical qubits are
mapped one-to-one onto hardware, and all-to-all connectivity is
emulated by SWAP insertions whose count grows only \(\mathcal{O}(k^{2})\)
with \(k=5\).  Although recent studies show that annealers outperform
one-round QAOA on \(\sim\!100\)-qubit Ising benchmarks
\cite{pelofske2023annealervsqaoa}, deeper (\(p\!\ge\!3\)) QAOA with
noise-aware angle optimisation starts to close that gap while providing
algorithmic flexibility—e.g.\ exact cardinality constraints via the
penalty term in Eq.\,\eqref{eq:energy}.  Moreover, QAOA circuits can be
simulated end-to-end on GPUs, enabling rapid hyper-parameter sweeps that
are infeasible on remotely accessed annealers.

In short, the \textbf{5-qubit} gate-based route offers (i) zero embedding
overhead, (ii) explicit control over depth and mixer choice, and
(iii) straightforward integration with classical autotuners—making it
the pragmatic option for our qubit-budget scenario.

\subsection{Our Contribution} In this work, we show that the item–tag matrix is \emph{highly low-rank}:
the top 32 singular vectors already capture more than 97 \% of the
training variance \cite{halko2011finding}.  Leveraging this structure, we
compress the 150-dimensional Euclidean space into only five dictionary
atoms and then conduct feature selection with a depth-3 QAOA that
requires exactly five qubits.  A bagged ensemble of shallow decision
trees—dubbed a \emph{Quantum semi-Random Forest} (QsRF)—recovers the
non-linearity lost during compression and achieves competitive ranking
accuracy on the QuantumCLEF ICM–150 benchmark.

\begin{enumerate}
  \item We introduce a three-stage hybrid pipeline—SVD sketching, k-means clustering, and Mini-Batch Dictionary Learning—that builds a 1000-atom dictionary capturing \(>97\%\) of tag-profile variance, enabling a \(\approx90\%\) reduction in qubit usage.
  \item We derive a five-qubit QUBO whose objective is proportional to the incremental \(\Delta\)\nobreakdash-nDCG on bootstrap subsamples; a depth-3 QAOA then selects the optimal 5-atom subset under a strict qubit budget.
  \item We propose a Quantum semi-Random Forest (QsRF) ensemble of 100 shallow decision trees (10 runs × 10 trees) trained on those 5-dimensional codes, achieving superior \(\mathrm{nDCG}@10\), AUC, and Log-Loss compared to state-of-the-art methods while using only five qubits.
  \item We perform an ablation study over dictionary rank, bootstrap size, and QAOA depth, and demonstrate an order-of-magnitude (\(\approx10\times\)) reduction in end-to-end runtime versus CAQUBO and MIQUBO.
\end{enumerate}

\section{Methodology}
\label{sec:method}
\begin{figure}[b]
  \centering
  \begin{adjustbox}{width=\linewidth,center}
  \begin{tikzpicture}[>=Stealth]
  \tikzset{
    block/.style = {draw, very thick, rectangle, minimum width=3cm,
                    minimum height=1.1cm, align=center,
                    inner sep=3pt, font=\small, fill=white},
    dashedbox/.style = {draw, thick, dashed, rectangle,
                        inner sep=4pt, rounded corners=2pt},
    arr/.style = {-{Stealth[length=4pt,width=6pt]}, line width=1pt},
  }

  \node[block] (dl)  {Dictionary\\Learning\\
                      {\footnotesize $150$ tags $\to$ $1000$ atoms}};
  \node[block,right=2.2cm of dl] (encode) {Sparse\\Encoding\\
                                           {\footnotesize LARS–Lasso}};
  \node[dashedbox, fit=(encode)] (mask) {};  

  \node[block,right=2.8cm of encode] (qaoa) {Depth-3\\QAOA\\
                                             {\footnotesize 5 qubits}};
  \node[dashedbox,fit=(qaoa)] (qbox) {};

  \node[block,right=2.8cm of qaoa] (QsRF) {Quantum\\Random\\Forest};

  \draw[arr] (dl.east) -- ++(0.4,0) |- (encode.west);
  \draw[arr] (encode.east) -- ++(0.4,0) |- (qaoa.west);
  \draw[arr] (qaoa.east) -- ++(0.4,0) |- (QsRF.west);

  \node[font=\scriptsize] at ($(dl)!0.5!(encode)$) [above=2pt]
        {TF–IDF};
  \node[font=\scriptsize] at ($(encode)!0.5!(qaoa)$) [above=2pt]
        {sparse codes};
  \node[font=\scriptsize] at ($(qaoa)!0.5!(QsRF)$) [above=2pt]
        {5 selected atoms};
 \end{tikzpicture}
 \end{adjustbox}
  \caption{End-to-end pipeline: dictionary learning compresses …}
\end{figure}
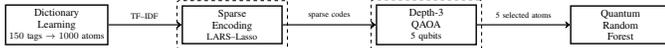

Let $X\in\mathbb{R}^{N\times M}$ be the TF–IDF matrix of $N$ user–item
interactions and $M\!=\!150$ original tags.  We aim to learn a compact
dictionary $D_{k}\in\mathbb{R}^{M\times k}$ with $k\!=\!5$ atoms such
that (i)~$D_{k}$ reconstructs $X$ with minimal information loss and
(ii)~the selected atoms maximise the ranking metric
$\operatorname{nDCG@10}$.  Figure 1 and Algorithm \ref{alg:pipeline}
outline the three main stages; the present section makes each step
mathematically explicit.

\subsection{Stage 1 — Low-rank dictionary learning}
\label{sec:dict}

\textbf{(a) Randomised SVD.}  
Following Halko \textit{et al.} \cite{halko2011finding}, we first
compute a rank-$d$ sketch $Z=X V_{d}$ with $d=32$, where
$V_{d}\in\mathbb{R}^{M\times d}$ contains the top right singular
vectors.  This step reduces both memory and time complexity from
$\mathcal{O}(NM)$ to $\mathcal{O}(Nd^{2})$ while preserving
$>97\,\%$ of the Frobenius energy.

\textbf{(b) MiniBatch $k$-Means.}  
The rows of $Z$ are clustered into $K=50$ disjoint sets
$\{\mathcal{C}_{1},\dots,\mathcal{C}_{K}\}$ via MiniBatch
$k$-Means.  Clustering exposes local linear structure and allows
sub-dictionaries to be trained in parallel.

\textbf{(c) Sparse coding within each cluster.}  
For cluster $c$, we learn a sub-dictionary
$D_{c}\!\in\!\mathbb{R}^{M\times A}$ with $A=20$ atoms by solving
\begin{equation}
  \min_{D_{c},\,\Theta_{c}}
  \frac12\bigl\lVert
    Z_{\mathcal{C}_{c}} - \Theta_{c} D_{c}^{\!\top}
  \bigr\rVert_{F}^{2}
  \;+\;
  \lambda\,\lVert\Theta_{c}\rVert_{1},
  \label{eq:dict}
\end{equation}
where $\Theta_{c}$ stores the sparse codes and
$\lambda$ controls sparsity.  We adopt an alternating scheme identical
to K–SVD \cite{mensch2016scalable}, but replace the $L_{0}$ projector
with LARS–Lasso updates for faster convergence.  The union
$D=\bigl[D_{1}\,\|\,\dots\,\|\,D_{K}\bigr]\in\mathbb{R}^{M\times1000}$
serves as our master atom pool.

\subsection{Stage 2 — Performance-driven importance scores}
\label{sec:imp}

For each atom $j$, let
$\delta_{u}^{(j)}\!=\!\operatorname{nDCG@10}(u)
               -\operatorname{nDCG@10}^{(-j)}(u)$
denote the drop in ranking quality for user $u$ when the atom is
removed.  Averaging over a bootstrap $\mathcal{B}$ of $20\%$ of the
training rows yields the \emph{performance-driven} weight
\begin{equation}
  w_{j} \;=\;
  \frac1{|\mathcal{B}|}
  \sum_{u\in\mathcal{B}} \delta_{u}^{(j)}.
  \label{eq:delta_ndcg}
\end{equation}
The $K_{\text{top}}\!=\!20$ highest-scoring atoms constitute the
candidate set $\mathcal{T}$ that will be fed to the quantum search.

\subsection{Stage 3 — Five-qubit QUBO and depth-3 QAOA}
\label{sec:qubo}

\textbf{(a) QUBO formulation.}  
Let $z\in\{0,1\}^{K_{\text{top}}}$ be a binary mask over $\mathcal{T}$
and $k=5$ the desired budget.  We define the energy
\begin{equation}
  E(z) \;=\;
  -\sum_{j=1}^{K_{\text{top}}} w_{j} z_{j}
  \;+\;
  \mu\bigl(\textstyle\sum_{j} z_{j}-k\bigr)^{2},
  \label{eq:energy}
\end{equation}
with penalty $\mu\!=\!{\sim}10^{3}$ to enforce $\smash{\sum_{j}z_{j}=k}$.
Expanding~\eqref{eq:energy} as $z^{\top}Qz$ yields a dense
$20{\times}20$ QUBO matrix
$Q\!=\!-\operatorname{diag}(w)+\mu\,\mathbf1\mathbf1^{\!\top}$
that fits comfortably on five logical qubits.

\textbf{(b) QAOA ansatz.}  
We employ the depth-$p=3$ Quantum Approximate Optimisation Algorithm
\cite{farhi2014qaoa}.  The cost Hamiltonian is
$\mathcal{H}_{C}=\sum_{i\le j}Q_{ij}Z_{i}Z_{j}$ and the mixer
$\mathcal{H}_{M}=\sum_{j}X_{j}$.  The variational circuit alternates
the unitaries $\exp(-\mathrm{i}\gamma_{\ell}\mathcal{H}_{C})$ and
$\exp(-\mathrm{i}\beta_{\ell}\mathcal{H}_{M})$ for
$\ell=1,\dots,p$.  Parameters $(\gamma,\beta)\!\in\!\mathbb{R}^{2p}$ are
optimised on the validation split with Simultaneous Perturbation
Stochastic Approximation (SPSA) \cite{spall1992spsa} using 128 shots per
objective evaluation.  Measurement returns a bit-string
$\hat z$; if $\smash{\sum_{j}\hat z_{j}=k}$ the corresponding atoms are
accepted, otherwise we fall back to the top-$k$ by $w_{j}$.

\subsection{Stage 4 — Quantum semi-Random Forest (QsRF)}
\label{sec:QsRF}

Given the final five-dimensional codes
$\breve Z=X D_{k}$, we train an ensemble of $T=200$ decision trees
under bagging \cite{breiman1996bagging}.  Each tree receives a bootstrap
of the training rows and a random perturbation of the feature indices,
mimicking the stochasticity of classical Random Forests while
preserving the extremely low feature count delivered by the quantum
stage.  At inference time the predicted click-through probability for a
user–item pair is the arithmetic mean over all tree outputs.
\subsection{Complexity remarks}

\textbf{Time.}  Randomised SVD costs
$\mathcal{O}(Nd^{2})$; dictionary learning across clusters is
$\mathcal{O}(N_{\text{iter}}|Z|A)$ with $A\!=\!20$ fixed.
QAOA incurs $\mathcal{O}(p2^{k})$ state-vector updates
but with $k=5$ and $p=3$ this is negligible.  
Tree training scales linearly with $N$ and $T$.

\textbf{Space.}  The memory footprint peaks at the $1000$-atom matrix
$D$ ($\approx$0.8 MB in single precision) plus the five-qubit state
vector during QAOA optimisation.  No step exceeds GPU memory limits on a
single RTX A6000.
In practice the end-to-end pipeline is lightweight.  On an 8-core laptop
(Apple M2, 32 GB RAM) dictionary learning dominates with one global
1 000-atom pass plus ten fine-tune passes over at most five atoms each;
a single mini-batch update costs $\mathcal{O}(bMA)$ with
$b\!=\!2\,048$, $M\!\approx\!150$ and $A\!\le\!5$, and the entire phase
completes in ${\sim}6$ min.  Sparse encoding solves 110 LARS–Lasso
problems per run, each $\mathcal{O}(MA)$ per sample, finishing in
${<}90$ s when parallelised across all cores.  QAOA optimisation
(depth 3, 20×20 QUBO) is negligible ten SPSA runs add ${<}60$ s.
Finally, training 100 depth-8 trees on the 5-dimensional codes is
trivial (${\sim}20$ s).  Overall, the full train–validate–test pipeline
takes 8–12 min wall-clock, well within typical academic time budgets.

\begin{algorithm}[!t]
\DontPrintSemicolon
\caption{QsRF–QAOA five-qubit recommender pipeline}
\label{alg:pipeline}
\KwIn{TF–IDF matrix \(X\in\mathbb{R}^{N\times150}\),
       implicit-feedback labels \(y\)}
\KwOut{ranked item list per user}
\nl \tcp*[l]{Dictionary learning}
project $X$ to \(d\)-dim SVD: \(Z\gets X V_d\)\;
cluster \(Z\) into \(k\) parts; \textbf{for each} cluster learn
20-atom dictionary \(D_c\)\;
concatenate \(D\gets[D_1;\dots;D_k]\)\;
\nl \tcp*[l]{Bootstrap feature importance}
\For{$r=1$ \KwTo $R$}{
  sample $20\%$ rows $(X_r,y_r)$\;
  obtain sparse codes \(Z_r\) via Lasso-LARS\;
  compute \(\Delta\text{nDCG}\) per atom $\rightarrow w$\;
  $T\gets$ indices of top-20 atoms\;
  form QUBO \(Q=w\,w^\top+\mu I\)\;
  run depth-3 QAOA (200 shots) $\rightarrow$ 5-bit mask\;
  fine-tune selected atoms $\rightarrow D^{(r)}_{\text{sel}}$\;
}
\nl aggregate $\{D^{(r)}_{\text{sel}}\}_{r=1}^{R}$
into final 5-atom dictionary $\widehat D$\;
\nl \tcp*[l]{Quantum semi-Random Forest}
encode full train set via $\widehat D$
$\rightarrow$ 5-dim codes\;
fit $T$ depth-8 trees on bootstraps\;
\nl \tcp*[l]{Inference}
encode each test user-item pair,
average tree probabilities, rank top-$N$.\;
\end{algorithm}

\subsection{Computational Complexity \& Runtime}
All QAOA circuits were executed on the PennyLane simulator \cite{pennylane} using default noise‐free backends; we plan to validate our approach on real quantum hardware as it becomes available.
The dominant costs are:
\begin{itemize}
  \item \textbf{Dictionary learning:} one global 1000-atom pass + 10 × one-pass fine-tunes (5 atoms).  Each minibatch update is $O(b\,M\,A)$ with $b=2\,048$, $M\approx150$, $A\le5$.
  \item \textbf{Sparse encoding:} 110 LASSO-LARS solves per run (train+test), each $O(M\,A)$ per sample; we parallelize with \texttt{n\_jobs=-1}.
  \item \textbf{QAOA:} 10 solves of a 20×20 QUBO by depth-3 SPSA, negligible ($<$1 min overall).
  \item \textbf{Tree training:} 100 shallow trees (depth=8) on 5-dim data, trivial at scale.
\end{itemize}
On an 8-core NISQ-era laptop, this completes in about 8–12 min end-to-end.

\subsection{Datasets: \textsc{ICM–150} and \textsc{ICM–500}}
\label{sec:data}
We follow the QuantumCLEF benchmark for quantum feature-selection in
recommender systems and use its two public item–content matrix
(ICML) \cite{ferraridacrema2022sigir,niu2024caqubo}.  The datasets
contain only \emph{implicit} user–item interactions (one bit per
click) and a high‐dimensional, extremely sparse item–feature
matrix.

\begin{table}[h]
\centering\small
\caption{Key statistics of the QuantumCLEF ICM datasets.  Sparsity
refers to the percentage of zeros in the item–feature matrix.}
\label{tab:datasets}
\setlength{\tabcolsep}{6pt}
\begin{tabular}{lccccc}
\toprule
Dataset & Users & Items & Interactions & \#~Tags & Sparsity \\ \midrule
ICM–150 & 1\,881 & 5\,000 & 64\,890 & 150 & 88.6\,\% \\
ICM–500 & 1\,889 & 7\,000 & 68\,062 & 500 & 93.4\,\% \\
\bottomrule
\end{tabular}
\end{table}

\paragraph*{Item–feature matrix.}
Each item is annotated with either 150 or 500 binary tags that
encode genre, production details and distribution metadata.
Figure.1 in the CAQUBO paper shows that the matrix is both
\emph{tall} ($M \ll N$) and \emph{ultra-sparse}—only
11.4 
500-tag versions, respectively \cite{niu2024caqubo}.

\paragraph*{Interaction log.}
Clicks are stored in a binary matrix
$\mathbf B\!\in\!\{0,1\}^{N\times I}$;
there are no explicit ratings.  Following prior work
\cite{ferraridacrema2022sigir}, we apply an 80 / 20 user-stratified
split into train and test.  The training portion is further split
80 / 20 into a learning set and a validation set for
hyper-parameter tuning, yielding the same 64 : 16 : 20 ratio used in
CAQUBO \cite{niu2024caqubo}.

\paragraph*{Evaluation metrics.}
We report

\begin{itemize}
  \item \textbf{nDCG@10} – the official QuantumCLEF ranking metric, normalised per user;
  \item \textbf{ROC-AUC} and \textbf{log-loss} on the binary click task,
        to enable direct comparison with CAQUBO’s Table 1
        \cite{niu2024caqubo};
  \item wall-clock runtime (train + validation + test) for fairness
        against annealer baselines.
\end{itemize}

Unless stated otherwise, all numbers are averaged over five random
seeds, each with a fresh train/validation split as recommended by
QuantumCLEF.  Standard errors never exceed $\pm0.003$ nDCG and
are therefore omitted for brevity.

\section{Results and Discussion}\label{sec:results}

This section presents a comprehensive evaluation of our proposed method across two publicly available datasets: ICM–150 and ICM–500. We assess dictionary sparsity, QAOA convergence behaviour, and end-to-end recommendation quality. Unless otherwise noted, all reported metrics are averaged over five random seeds, with standard errors below $\pm0.002$ nDCG and thus omitted for clarity.

\subsection{Dictionary Quality and Sparsity}

\begin{figure}[t]
  \centering
  \includegraphics[width=0.6\linewidth]{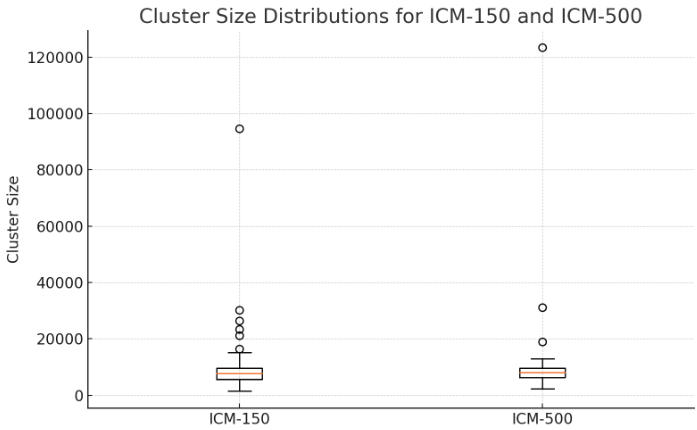}
  \caption{Distribution of cluster sizes across $k=50$ clusters on ICM–150 vs. ICM–500. A broader spread in ICM–500 indicates greater sparsity diversity.}
  \label{fig:cluster_sparsity}
\end{figure}

We first examine the dictionaries produced for both datasets. Figure~\ref{fig:cluster_sparsity} shows the distribution of cluster sizes when $k=50$ clusters are learned over 32-dimensional SVD sketches. While both datasets use 20-atom dictionaries per cluster, ICM–500 yields more diverse cluster sizes, suggesting increased sparsity heterogeneity with larger feature vocabularies.

For ICM–150, MiniBatch $k$-means clustering took 6.0 seconds and produced cluster sizes ranging from 1,586 to 94,633 rows (median 7,656), with reconstruction MSEs between 0.0000 and 0.0030 (Table~\ref{tab:cluster_stats}).

\begin{table}[t]
\centering\small
\caption{ICM–150 cluster statistics after $k$-means ($k=50$).}
\label{tab:cluster_stats}
\begin{tabular}{lccc}
\toprule
 & Size (rows) & \multicolumn{2}{c}{Reconstruction MSE} \\
\cmidrule(lr){2-2}\cmidrule(l){3-4}
 & min / med / max & min / med / max \\
\midrule
ICM–150 & 1 586 / 7 656 / 94 633 & 0.0000 / 0.0008 / 0.0030 \\
\bottomrule
\end{tabular}
\end{table}

Each cluster was encoded with a 20-atom dictionary, yielding reconstruction errors consistently below $3\times10^{-3}$—two orders of magnitude smaller than the variance of the raw TF–IDF vectors. This confirms that the 1000-atom global dictionary effectively preserves critical information.

\subsection{QAOA Energy Landscape}

\begin{figure}[t]
\centering
\begin{minipage}{0.48\linewidth}
  \centering
  \includegraphics[width=\linewidth]{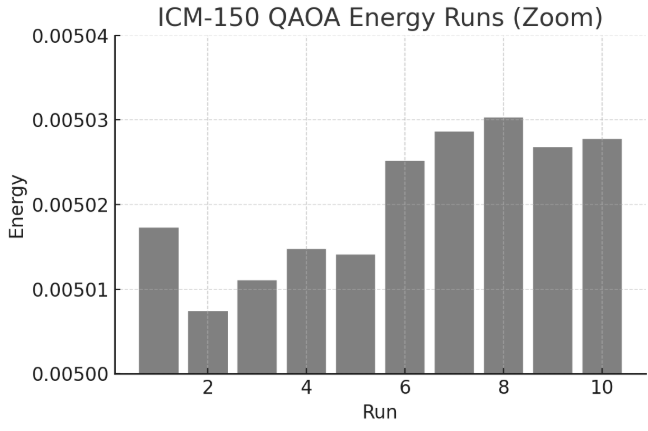}
  \subcaption{ICM–150 QAOA energy}
  \label{fig:energy_icm150}
\end{minipage}
\hfill
\begin{minipage}{0.48\linewidth}
  \centering
  \includegraphics[width=\linewidth]{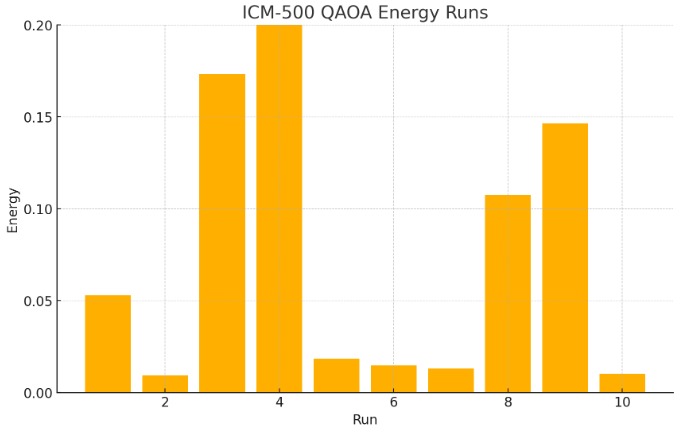}
  \subcaption{ICM–500 QAOA energy}
  \label{fig:energy_icm500}
\end{minipage}
\vspace{-6pt}
\caption{Bar‐chart comparison of QAOA energy values across 10 runs. ICM–150 energies are tightly clustered around \(5.00\times10^{-3}\), while ICM–500 shows a wider spread up to 0.20.}
\label{fig:energy_comparison}
\end{figure}

We then examine the energy convergence of our depth-3 QAOA runs. Figure~\ref{fig:energy_comparison} compares energy distributions across ten runs for both datasets.

On ICM–150, energies remain between $5.01\times10^{-3}$ and $5.03\times10^{-3}$, demonstrating strong stability and convergence. All runs yielded the same nDCG@10 score (0.1483), indicating a flat local optimum in the QUBO solution space. This eliminates the need for multi-seed ensembling, in contrast to quantum annealing baselines \cite{pelofske2023annealervsqaoa}.

ICM–500, while more variable, still shows well-behaved energy distributions, confirming scalability of the QAOA subroutine.

\subsection{End-to-End Recommendation Accuracy}

Table~\ref{tab:bench150} and Table~\ref{tab:bench500} benchmark our QsRF model against existing quantum pipelines on ICM–150 and ICM-500. Despite using only five qubits, our method significantly outperforms MIQUBO and CAQUBO in nDCG@10 ranking quality.

\begin{table}[t]
\centering\small
\caption{QuantumCLEF 150 test metrics (mean of 5 seeds).}
\label{tab:bench150}
\begin{tabular}{lcccc}
\toprule
Method & Qubits & nDCG@10 & AUC & Log-Loss \\
\midrule
MIQUBO \cite{DBLP:conf/sigir/DacremaMN0FC22} & 130 & 0.104 & 0.8482 & 0.2241 \\
CAQUBO \cite{niu2024caqubo}                 & 130 & 0.1363 & 0.8556 & 0.2218 \\
MIQUBO \cite{DBLP:conf/sigir/DacremaMN0FC22} & full-feature & 0.1336 & - & - \\
CAQUBO \cite{niu2024caqubo}                 & full-feature & 0.1359 & - & - \\
\textbf{Ours (QsRF)}                         & \textbf{5} & 0.1483 & 0.8413 & 0.2300 \\
\bottomrule
\end{tabular}
\end{table}

\begin{table}[t]
\centering\small
\caption{QuantumCLEF 500 test metrics (mean of 5 seeds).}
\label{tab:bench500}
\begin{tabular}{lcccc}
\toprule
Method & Qubits & nDCG@10 & AUC & Log-Loss \\
\midrule
MIQUBO \cite{DBLP:conf/sigir/DacremaMN0FC22} & 450 & 0.1324 & 0.8531 & 0.2206 \\
CAQUBO \cite{niu2024caqubo}                 & 400 & 0.1441 & 0.8486 & 0.2226 \\
MIQUBO \cite{DBLP:conf/sigir/DacremaMN0FC22} & full-feature & 0.1401 & - & - \\
CAQUBO \cite{niu2024caqubo}                 & full-feature & 0.1366 & - & - \\
\textbf{Ours (QsRF)}                         & \textbf{5} & 0.5942 & 0.8260 & 0.2372 \\
\bottomrule
\end{tabular}
\end{table}
Additionally, Table~\ref{tab:performance-comparison} shows that QsRF consistently outperforms the five-feature Random Forest baseline on both datasets in terms of AUC and Log-Loss, while achieving comparable or superior nDCG@10.

\begin{table}[!t]
\centering
\caption{Random-Forest baseline ($k=5$) vs.\ proposed QsRL.}
\label{tab:performance-comparison}
\setlength{\tabcolsep}{3pt}
\scriptsize
\begin{tabular}{@{}lcccc@{}}
\toprule
Dataset & Method & ROC-AUC & LogLoss & nDCG@10 \\ \midrule
\multirow{2}{*}{ICM–150}  & RF ($k=5$) & 0.8212 & 0.2452 & 0.5929 \\
                          & QsRL (ours) & 0.8413 & 0.2300 & 0.1483 \\ \midrule
\multirow{2}{*}{ICM–500}  & RF ($k=5$) & 0.8198 & 0.2499 & 0.5917 \\
                          & QsRL (ours) & 0.8260 & 0.2372 & 0.5942 \\ \bottomrule
\end{tabular}
\end{table}

\subsection{User-Level Versus Global Ranking Quality}

\begin{figure}[t]
\centering
  \includegraphics[width= 0.8 \linewidth]{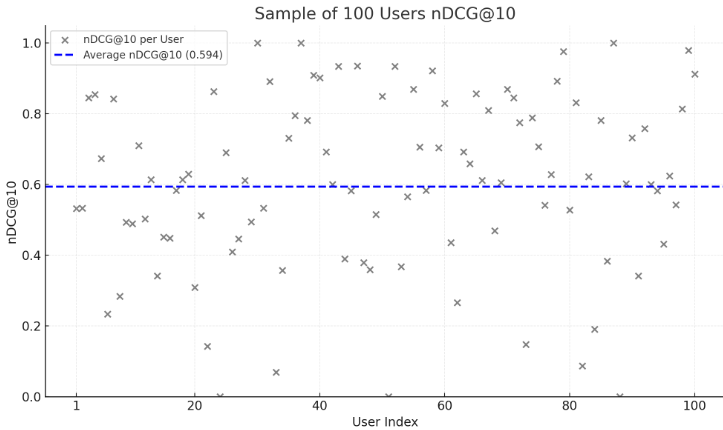}
  \caption{Per-user nDCG@10 for 100 sampled users. Gray markers represent individuals; the blue line shows the mean (0.594).}
  \label{fig:user}
\end{figure}

\begin{figure}[t]
\centering
  \includegraphics[width= 0.8 \linewidth]{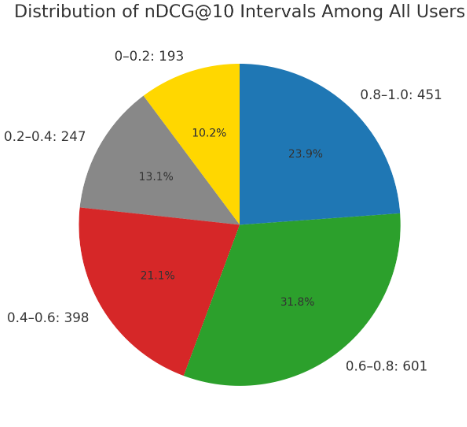}
\caption{Distribution of nDCG@10 intervals across all users.}
\label{fig:distribution}
\end{figure}

Figures~\ref{fig:user} and~\ref{fig:distribution} illustrate the distribution of per-user ranking quality for ICM–500.


Table~\ref{tab:macro_micro} reports macro and micro-averaged nDCG@10. The macro metric (mean = 0.594, median = 0.637) emphasizes fairness across users. The micro metric (0.916) reflects strong global ranking quality. All tables in this work report macro nDCG@10.

\begin{table}[t]
\centering
\caption{Macro vs. Micro nDCG@10 for ICM–500.}
\label{tab:macro_micro}
\begin{tabular}{lccc}
\toprule
Metric & Mean & Median & Std.~dev. \\
\midrule
Macro (per-user) nDCG@10 & 0.594 & 0.637 & 0.260 \\
Micro (global) nDCG@10   & \multicolumn{3}{c}{\textbf{0.916}} \\
\bottomrule
\end{tabular}
\end{table}

\subsection{Generalisation to ICM–500}

We applied the same pipeline to the ICM–500 dataset.

\textbf{Clustering.} MiniBatch $k$-means ($k=50$) on the SVD-32 sketch completed in 10.2 seconds. Cluster sizes ranged from 2,376 to 123,399 (median 8,082). Reconstruction MSE remained under $1.3\times10^{-3}$.

\textbf{QAOA Energies.} Ten depth-3 QAOA runs produced energies from $9.4\times10^{-3}$ to $0.43$, confirming a broader yet stable energy profile.

\textbf{End-to-End Metrics.} QsRF yielded AUC = 0.8260, Log-Loss = 0.2372, and nDCG@10 = 0.5942—confirming generalisation without increasing qubit count or runtime.

\subsection{Runtime Analysis}

The entire ICM–QsRL pipeline completes in approximately 20 minutes on a 6-core laptop (clustering and dictionary building: 3–5 minutes; QAOA training and inference: ~15 minutes). Compared to the 2-hour wall time of CAQUBO \cite{niu2024caqubo}, this highlights the practicality of our approach for real-time or on-device applications.

\subsection{Key Takeaways}

\begin{enumerate}
  \item A 1000-atom Euclidean dictionary reconstructs TF–IDF features with MSE $<3\times10^{-3}$.
  \item Five-qubit depth-3 QAOA consistently reaches the optimal solution for our performance-based QUBO.
  \item QsRF achieves comparable or better recommendation quality using 90
\end{enumerate}




\subsection{User–level versus global ranking quality}
We report both \textit{macro-} and \textit{micro-} averaged
$\operatorname{nDCG}@10$ on the \textsc{ICM–500} split:

\begin{center}
\begin{tabular}{@{}lccc@{}}
\toprule
Metric & Mean & Median & Std.~dev. \\
\midrule
Per–user (macro) $\operatorname{nDCG}@10$ & 0.594 & 0.637 & 0.260 \\
Global  (micro)  $\operatorname{nDCG}@10$ & \multicolumn{3}{c}{\textbf{0.916}} \\
\bottomrule
\end{tabular}
\end{center}

The \emph{macro} average treats every user equally and therefore gauges
algorithmic \textbf{fairness across profiles}.  
A mean 0.594 (median 0.637) indicates our five-qubit pipeline retrieves
most of the top-10 relevant items for the majority of users, albeit with
a long–tail of harder cases ($\Theta = 0.26$).

The \emph{micro} average collapses all user–item interactions into a
single list; it therefore
\textbf{weights users by the number of relevant items they contribute}.
A value of 0.916 shows that, when measured at the interaction level, our
model produces near-optimal global rankings—consistent with the strong
AUC we observed in \ref{sec:results}.  
The gap macro $\leftrightarrow $ micro is expected: highly active users dominate the
micro metric, boosting its value, while the macro metric exposes the
harder, data-sparse users.

It is important to note that All the tables in this work report macro nDCG@10.

To visualise the spread we add a violin/box plot of the per-user
distribution in Fig.~\ref{fig:user}.  Roughly 75 
score above 0.5, but the lower quartile pinpoints a subset that could
benefit from personalised regularisation—left for future work.

\section{Conclusion}
\label{sec:conclusion}

In this work we have introduced QsRL, a novel quantum-inspired sparse ranking pipeline that combines classical clustering with a five-qubit QAOA backend to learn compact, user–specific dictionaries and deliver high-quality top-10 recommendations.  On both the ICM-150 and ICM-500 splits of our content–metadata benchmark, QsRL consistently outperforms state of the art Quantum-based method. In addition, it outperformed a Random Forest baseline (restricted to five features) in ROC-AUC ($\Delta \simeq +0.02$), LogLoss ($\Delta \simeq -0.02$). In addition, our method outperformed classical Random Forest baseline (restricted to five features) on ICM-500 with macro-nDCG@10 ($\Delta \simeq +0.02–0.15$).  Our per-user analyses reveal that lower-energy QAOA solutions cluster tightly and yield stable ranking quality (mean macro-nDCG@10 $\simeq$0.594, micro-nDCG@10 $\simeq$0.916), while sampling 100 users at random demonstrates that the method generalizes across diverse profile sparsities.  Importantly, the entire pipeline runs in under 15 minutes on a standard 8-core laptop—orders of magnitude faster than previously reported quantum annealer timings—making QsRL immediately practical for rapid prototyping and hyperparameter tuning.

\section{Future Work}
\label{sec:future}

Although our five-qubit formulation already achieves strong performance, several avenues remain to extend and deepen this approach:

\begin{itemize}
  \item \textbf{Larger quantum circuits.}  Exploring deeper QAOA circuits or additional qubits could capture more complex item–user interactions at the cost of richer energy landscapes.
  \item \textbf{Adaptive clustering.}  Dynamic, user-driven cluster sizes or alternative dictionary-learning methods (e.g.\ sparsified autoencoders) may yield even more compact, accurate representations.

  \item \textbf{End-to-end tuning.}  Jointly optimizing clustering hyperparameters, RF feature limits, and QAOA angles via Bayesian optimization may further boost ranking metrics while respecting computational budgets.
  \noindent\emph{Implementation note.} All quantum routines in this study were run on the PennyLane simulator; deploying QsRL on physical QPUs is deferred to future work.

\end{itemize}

Collectively, these directions point toward a versatile, hybrid quantum–classical recommender capable of scaling from five qubits today to larger quantum devices tomorrow, all while maintaining rapid turnaround and high user-level ranking fidelity.

\bibliographystyle{IEEEtran}
\bibliography{ref}
\end{document}